\documentclass[twoside,11pt]{article}

\usepackage[accepted]{melba}

\usepackage{mwe} 

\usepackage{amsmath,amsfonts}

\melbaid{2023:010}  
\doi{https://doi.org/10.59275/j.melba.2023-1g8b}
\melbaauthors{Sina Amirrajab}  
\volume{2}
\firstpageno{288}  
\melbayear{2023}  
\datesubmitted{11/2022}  
\datepublished{06/2023}  

\ShortHeadings{Cardiac Pathology Synthesis}{Sina Amirrajab}

\usepackage{xcolor}
\title{Pathology Synthesis of 3D-Consistent Cardiac MR Images using 2D VAEs and GANs}

    \author{\name Sina Amirrajab  \email s.amirrajab@tue.nl \\  
	\addr Department of Biomedical Engineering, Eindhoven University of Technology, Eindhoven, The Netherlands.
	\AND
	\name Yasmina Al Khalil \email y.al.khalil@tue.nl \\ 
	\addr Department of Biomedical Engineering, Eindhoven University of Technology, Eindhoven, The Netherlands.
	\AND
	\name Cristian Lorenz  \email cristian.lorenz@philips.com \\ 
	\addr Philips Research Laboratories, Hamburg, Germany.
	\AND
	\name Jürgen Weese  \email juergen.weese@philips.com \\ 
	\addr Philips Research Laboratories, Hamburg, Germany.
	\AND
	\name Josien Pluim \email j.pluim@tue.nl \\ 
	\addr Department of Biomedical Engineering, Eindhoven University of Technology, Eindhoven, The Netherlands.
	\AND
	\name Marcel Breeuwer \email m.breeuwer@tue.nl \\ 
	\addr Department of Biomedical Engineering, Eindhoven University of Technology, Eindhoven, The Netherlands.
	\addr Philips Healthcare, MR R\&D - Clinical Science, Best, The Netherlands.
}

\begin{document}

\maketitle

\begin{abstract}

	We propose a method for synthesizing cardiac magnetic resonance (MR) images with plausible heart pathologies and realistic appearances for the purpose of generating labeled data for the application of supervised deep-learning (DL) training. The image synthesis consists of label deformation and label-to-image translation tasks. The former is achieved via latent space interpolation in a VAE model, while the latter is accomplished via a label-conditional GAN model. We devise three approaches for label manipulation in the latent space of the trained VAE model; i) \textbf{intra-subject synthesis} aiming to interpolate the intermediate slices of a subject to increase the through-plane resolution, ii) \textbf{inter-subject synthesis} aiming to interpolate the geometry and appearance of intermediate images between two dissimilar subjects acquired with different scanner vendors, and iii) \textbf{pathology synthesis} aiming to synthesize a series of pseudo-pathological synthetic subjects with characteristics of a desired heart disease. Furthermore, we propose to model the relationship between 2D slices in the latent space of the VAE prior to reconstruction for generating 3D-consistent subjects from stacking up 2D slice-by-slice generations. We demonstrate that such an approach could provide a solution to diversify and enrich an available database of cardiac MR images and to pave the way for the development of generalizable DL-based image analysis algorithms. We quantitatively evaluate the quality of the synthesized data in an augmentation scenario to achieve generalization and robustness to multi-vendor and multi-disease data for image segmentation. Our code is available at \url{https://github.com/sinaamirrajab/CardiacPathologySynthesis}.
\end{abstract}

\begin{keywords}
	Cardiac Pathology Synthesis, Image Synthesis, Conditional GANs, VAEs
\end{keywords}

\section{Introduction}

Deep generative modeling has gained attention in medical imaging research thanks to its ability to generate highly realistic images that may alleviate medical data scarcity \citep{GANsMedical1}. The most successful family of generative models known as generative adversarial networks (GANs) \citep{GANs} and Variational Autoencoders (VAEs) \citep{VAE} are widely used for medical image synthesis \citep{Joyce, GANreview}. Many studies have proposed generative models to synthesize realistic and diversified images for brain \citep{kwon2019generation, dar2019image, fernandez2022can} and heart \citep{Agis,rezaei2021generative} among other medical applications \citep{MedGAN}. Recently, diffusion models \citep{ho2020denoising} have been introduced as a new category of deep generative models. They aim to address some of the challenges related to GANs and VAEs, particularly regarding sampling diversity and quality. These models have shown promising results for synthesizing brain images \citep{pinaya2022brain}. However, the generated data are often unlabeled and therefore not suitable for training a supervised deep learning algorithm, for instance, for medical image segmentation.

Despite the benefit of data augmentation and anonymization using synthetic data for brain tumor segmentation \citep{shin2018medical,yu20183d}, the application of synthesizing labelled cardiac MRI data remained relatively under-explored with very limited recent attempts to synthesize cardiac images \citep{skandarani2021generative}. Recent work by \citep{skandarani2020effectiveness} and \citep{MICCAI2020, amirrajab2022label} investigate the effectiveness of using conditional GANs for translating ground truth labels to realistic cardiac MR images that do not require manual segmentation and can be used for training a supervised segmentation model. However, the images are generated on a fixed set of input labels and therefore create similar heart anatomies, very limited to available ground truth labels of the training data, and more importantly, unable to synthesize subjects with cardiac pathology.

\section{Contribution}

\begin{figure}[!ht]
    \centering
    \includegraphics[width=0.95 \linewidth]{ 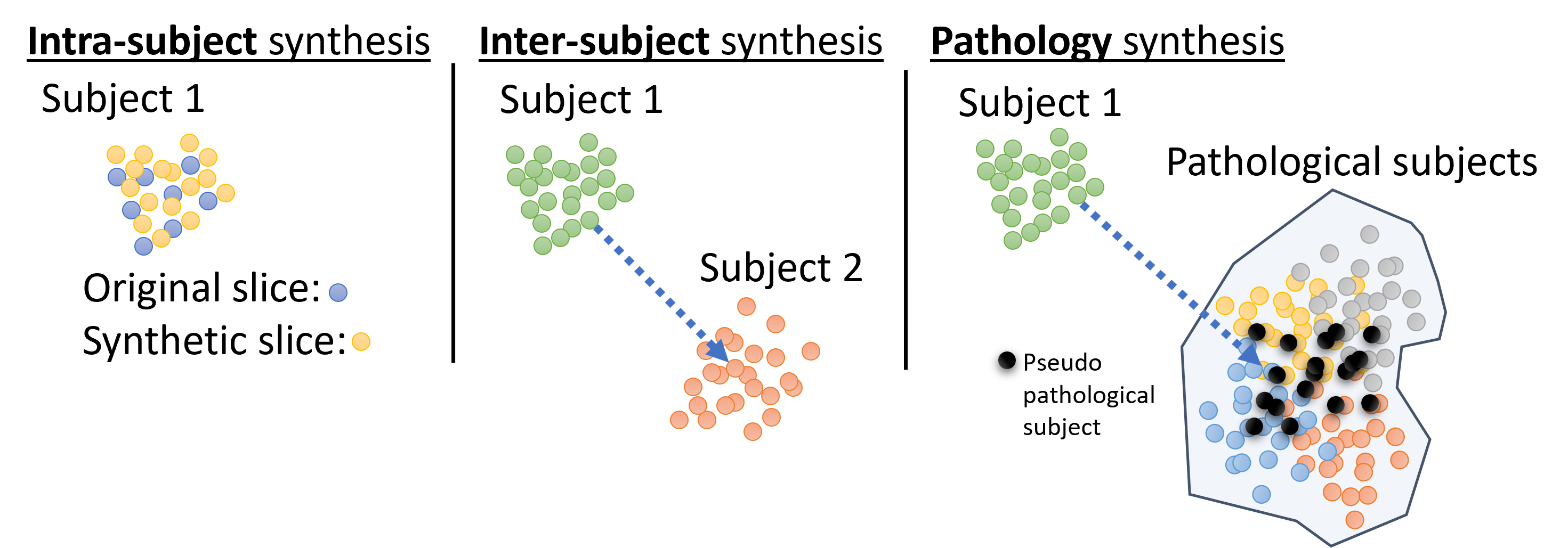}
    \caption{Three strategies to traverse and interpolate in the latent space to perform label deformation using the trained VAE model. Each encoded slice of a subject is represented as a dot in the low-dimensional latent space. The number of slices is increased using cubic interpolation in the latent space for intra-subject synthesis, and intermediate latent codes between two subjects (Subject 1 and 2) are generated using linear interpolation for inter-subject synthesis, indicated as dotted blue arrows. Assuming that all pathological subjects can be clustered in a neighboring location of the latent space, the statistics are estimated to draw a sample (pseudo-pathological subject) for pathology synthesis. Interpreting  between Subject 1 and the pseudo-pathological subject results in generating subjects with pathological characteristics.}
    \label{fig:my_label_1}
\end{figure}

We propose to break down the task of cardiac image synthesis into 1) learning the deformation of anatomical content of the ground truth (GT) labels using VAEs and 2) translating GT labels to realistic cardiac magnetic resonance (MR) images using conditional GANs. We generate various virtual subjects via three approaches, namely i) \textbf{intra-subject synthesis} to improve the through-plane resolution and generate intermediate short-axis slices within a given subject, ii) \textbf{inter-subject synthesis} to generate intermediate heart geometries and appearance between two dissimilar subjects scanned using two different scanner vendors, and iii) \textbf{pathology synthesis} to generate virtual subjects with a target heart disease that affects the heart geometry, e.g. synthesizing a pseudo-pathological subject with thickened myocardium for hypertrophic cardiomyopathy. All three approaches are accomplished via manipulation and interpolation in the latent space of our VAE model trained on GT labels, as demonstrated in Figure \ref{fig:my_label_1}. The synthetic subjects in this study are labeled by design and therefore suitable for medical data augmentation.

Furthermore, we propose a method to generate 3D consistent volumes of synthetic subjects by modelling the correlation between 2D slices in the latent space. The relationship between the slices is captured via estimating the covariance matrix calculated for all latent vectors across all slices. The estimated covariance matrix is used to correlate the elements of a randomly drawn sample in the latent space just before feeding it to the decoder part of the VAE. This technique results in a coherent sampling from the latent space and in turn reconstruction of more consistent 3D volume by stacking 2D slices generated from the 2D model. The intensity information is extracted from the style encoder part of the generator, and a style image is used for each 2D slice to maintain the consistency of pixel intensity during synthesis.

The presented work is a substantially extended version of \cite{SASHIMI} where:
\begin{itemize}
    \item In addition to pathology synthesis, we explore new ways of interpolation in the latent space of the proposed variational auto-encoder and provide more details about the strategies to generate new subjects. We provide new experiments and more visualizations to explain our methodology in more detail.
    \item We propose “inter-subject synthesis” to generate subjects with varying heart shapes that can resemble the changing heart shapes and appearance between two dissimilar real subjects scanned using two different scanner vendors. The aim of this approach is reducing the domain gap between data from two different sources.
    \item We take advantage of new publicly available cardiac MRI data to evaluate our model’s ability to generate new subjects using our proposed strategies, especially multi-vendor and multi-disease data from M\&Ms-1 \citep{MMs} and M\&Ms-2 \footnote{(https://www.ub.edu/mnms-2/)} challenges. 
    \item We quantitatively evaluate the usability of such generated subjects in a data augmentation scenario for training a cardiac MRI segmentation model. Our experiments show that augmentation helps the model achieve better generalization and robustness to multi-vendor and multi-disease data, particularly when we add synthetic images with pathology.
\end{itemize}

\section{Methods}

\subsection{Image Synthesis Model}

The synthesis model architecture includes a ResNet encoder \citep{ResNet} for extracting the style of an input image and a label conditional decoder based on Spatially Adaptive DE-normalization (SPADE) layers \citep{SPADE}. The SPADE layers preserve the anatomical content of the GT labels. After successful training of the model with pairs of real images and corresponding labels, the generator can translate GT labels to realistic cardiac MR images. To alter the heart anatomy of the synthesized image, we can simply deform the labels. In the previous studies, new subjects were synthesized by applying simple transformations such as random elastic deformation, morphological dilation, and erosion on GT labels \citep{lustermans2021optimized, al2021late}. The authors have demonstrated that the synthetic data generated using this approach, despite generating unrealistic anatomies, boosted the performance of medical image analysis models for tissue segmentation in cine and LGE cardiac MRI data. We utilize the same synthesis network with default training parameters for this study and here we propose a deep learning (DL) label deformation model based on Variational Autoencoders (VAEs).

\subsection{Label Deformation Model}
\label{label deformation}

We propose a DL-based approach using a VAE model to generate plausible anatomical deformations via latent space manipulation to generate subjects with characteristics of heart pathologies. The VAE model consists of an encoder and a decoder network trained on the ground-truth label masks and tries to learn the underlying geometrical characteristics of the heart present in the labels. The changes in heart geometry can be associated with a specific type of disease. For instance, thickening and thinning of the left ventricular myocardium can be an indicating factor of hypertrophic and dilated cardiomyopathy, respectively. The goal here is to learn the effects of these factors on the heart geometry presented in the GT labels and to explore the latent space of the VAE to generate new labels with plausibly deformed anatomies. Additionally, we model the characteristics of particular heart diseases in the latent space and generate new samples with heart geometries that represent these disease characteristics.

A convolutional VAE model is designed and trained on GT labels of the heart to learn the underlying factors of different heart geometries presented in the database. After training, we encode all data into the latent space using the encoder part of the VAE and perform different operations to manipulate the learned features of the data, in this case the heart geometry. For instance, once labels are encoded into the latent space, we can traverse between two locations by simply interpolating between two latent codes and performing reconstruction to generate new heart geometries with intermediate anatomical shapes. These newly deformed labels are used as an input of the label-conditional GANs for image synthesis.

\subsection{Approaches to Generate New Subjects}

We investigate approaches to generating new subjects using our trained VAE model, which involves 
deforming labels in an anatomically plausible manner via modifying the latent representation of the ground truth labels. A schematic view of the latent space and three distinct approaches to achieve this goal is illustrated in Figure \ref{fig:my_label_1}. Firstly, we define \textbf{Intra-subject} synthesis as an essential first step to increase the number of short-axis slices per subject and equalize that for all subjects in the dataset. This is accomplished by interpolating between the latent codes of different slices belonging to the same subject. Secondly, we propose \textbf{Inter-subject} synthesis to create synthetic subjects with heart geometry and imaging characteristic that are a combination of two different-looking subjects acquired using scanners from two different vendors. The intermediate subjects are interpolated between subject 1 and subject 2 in Figure \ref{fig:my_label_1} after intra-subject synthesis. Finally, we introduce \textbf{Pathology} synthesis to generate pseudo-pathological conditions in a normal subject. This approach enables us to explore the progression of heart disease and its potential impact on the heart's geometry. Overall, the second and third approaches aim to generate a diverse range of synthetic subjects, which can be useful in several medical applications, such as training and validating deep learning models for image segmentation.

\subsubsection{Intra- and Inter-subject Synthesis}

For intra-subject synthesis, we wish to increase the through-plane resolution of the short-axis slices for a subject. All slices (ranging between 6-13 slices per subject) are first encoded into the corresponding latent vectors. The latent vectors are then augmented by cubic interpolation to increase to 32 latent vectors, each representing one slice, which are then reconstructed by the decoder network to create labels for all 32 slices. All subjects will consist of 32 slices after the intra-subject synthesis. Note that the first and last slices are kept and only intermediate slices are interpolated.

Inter-subject synthesis aims to generate new examples with intermediate heart anatomy and appearance between two dissimilar subjects. To this end, intra-subject synthesis is first performed to equalize the number of short-axis slices for each subject in the latent space. Therefore, each encoded subject has 32 latent vectors associated with 32 interpolated slices. Then following the same procedure, intermediate latent vectors associated with in-between heart geometries are created by linear interpolation between the 32 latent vectors of the two encoded subjects. By decoding these newly interpolated subjects using the decoder part of the VAE, the heart geometry of one subject is morphed into the other one. Finally, the deformed labels are fed to the synthesis model for synthesizing new subjects. 

\begin{figure}[!ht]
    \centering
    \includegraphics[width=0.95\linewidth]{ 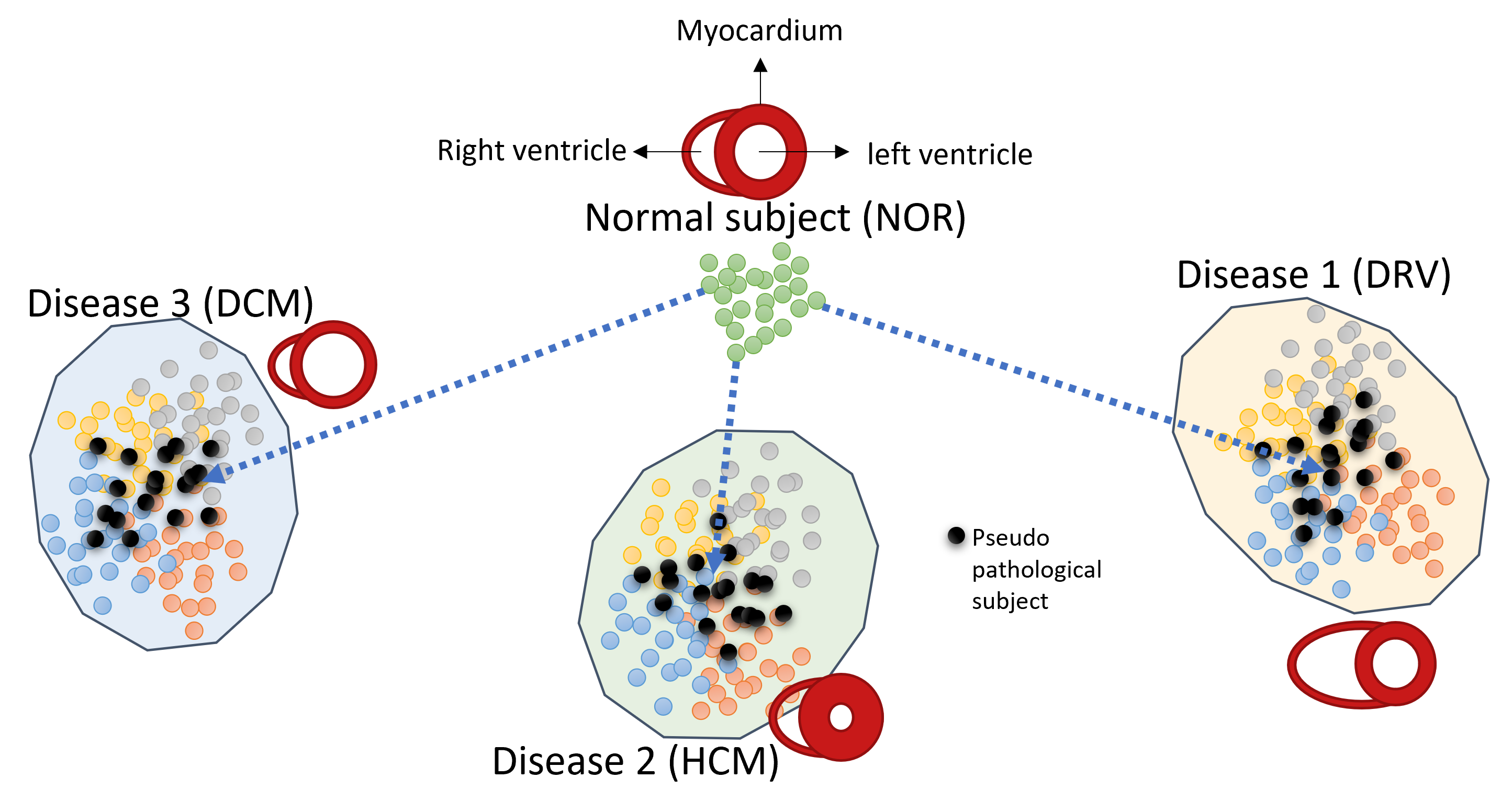}
    \caption{Pathology synthesis to generate a normal subject (NOR) with a target pathology such as dilated cardiomyopathy (DCM), hypertrophic cardiomyopathy (HCM) and dilated right ventricle (DRV), assuming that these diseases are clustered in the latent space.}
    \label{fig:my_label_4}
\end{figure}

\subsubsection{Pathology Synthesis}

Pathology synthesis is designed to generate subjects with informed characteristics of a heart pathology and its effects on the geometry of the heart, given that the pathology is manifested in the ground truth labels. The assumption here is that subjects with a common pathological class have similar heart characteristics and hence they are encoded to the same area in the latent space of the VAE trained with them. Figure \ref{fig:my_label_4} depicts a schematic illustration of the latent space information of a normal subject (with a normal heart shape) and a number of subjects with the same heart disease, having abnormal heart shapes, grouped in the same location. Each circle in the figure graphically represents the corresponding latent vector of an encoded slice for one subject.

Suppose we wish to generate subjects with a target pathology, for instance with characteristics of hypertrophic cardiomyopathy (HCM), potentially thickening of the myocardium. Note that we want to preserve the identity of a normal subject (NOR) and only generate disease characteristics such as thickening of the left myocardium for HCM. To this end, assuming that the disease features can be grouped to a neighboring location in the latent space, we encode all subjects with the desired pathology into the latent space and estimate mean, standard deviation, minimum, and maximum across all subjects for all interpolated slices; $[(\mu,\sigma,min,max)]_{HCM}$. These statistics are calculated on the mean of the latent vectors corresponding to the ground truth labels for subjects that are fed to the encoder network. The matrix size for these parameters is ($n_s\times n_z$), where $n_s$ is the number of interpolated slices (32 in our case) and $n_z$ is the size of the latent vector. Note that we equalize the number of slices for each subject via slice interpolation in the latent space. A sample is drawn from a truncated normal distribution parameterized by these statistics, which we call pseudo-pathology sample; $x_{pHCM}{\sim}TN[(\mu,\sigma, min, max)]_{HCM}$. The sample generated with statistics of all HCM subjects should potentially represent the heart features of a HCM subject: abnormally thick myocardium. We found that performing linear interpolation between a NOR subject and a pseudo-HCM sample gradually introduces pathological features of HCM to NOR subject's anatomy. Rather than simply reconstructing the pseudo-pathology sample, the interpolation process led to a mix of both normal and pathological characteristics in the reconstructed label. The patient's anatomy, in terms of the background area which includes organs other than the heart, is generated according to the style image provided to the image generator network.


\subsubsection{Modelling 3D Consistency}

The elements of the drawn sample in the latent space are assumed to be independent and identically distributed. To model the dependency of variables, the correlation between the dimensions of the latent code for all pathological subjects is measured using the Kendall rank correlation coefficient. The uncorrelated generated sample is then transformed in the latent space according to the overall correlation coefficient ($n_z\times n_z$) estimated from the training data to account for the relationship between elements of the latent code. The elements of the latent vector are correlated using Cholesky matrix decomposition as explained in the supplementary material. However, the relationship between different slices of one subject has not yet been modelled. This can lead to inconsistent heart geometries in the slice direction as a consequence of slice-by-slice 2D synthesis.

We propose a simple statistical approach to account for the relationship between slices in the latent space. During training, we train the VAE model using the ground truth labels of the 2D slices in the standard manner as described in Section \ref{label deformation}. However, during the inference time, we exploited the correlation between slices of an encoded subject in the latent space to improve the consistency between subsequent slices during synthesis. In pathology synthesis, we want to perform a linear interpolation between a NOR subject ($x_{NOR}$) and a random pseudo-pathological sample ($x_{pHCM}$). Although different slices of the NOR subject are inherently correlated in the latent space, the random sample does not contain any information about the relationship between slices. To model this relationship, we estimate the correlation between slices of the $x_{NOR}$  and construct the associated correlation coefficient matrix ($n_s \times n_s$). Given this matrix, we correlate the slices of the $x_{pHCM}$ using the Cholesky matrix decomposition. We model the relationship between slices during inference. We start by encoding all 2D slices of a subject into a latent space and use intra-subject synthesis to increase the number of slices to 32. This results in an interpolated subject with a vector size of 32 x nz, where nz is the size of the latent vector, which is 16 in our case. Next, we use the Kendall rank correlation coefficient to compute a correlation matrix of size 32 x 32, which measures the relationship between the encoded slices in the latent space. We decompose this matrix into a lower and upper triangular matrix using Cholesky decomposition. We then multiply the upper triangular matrix with a randomly generated sample in the latent space, which has uncorrelated slices, to create a new vector with the correlation between slices. Finally, we use the decoder to reconstruct the new vector and generate the corresponding labels. This procedure is explained in more detail in the supplementary material.

The interaction between elements of latent vectors as well as the relationship between different slices is modelled to generate more realistic correlated samples in the latent space. We found that both latent correlation matrix ($n_z\times n_z$) and slice correlation matrix ($n_s\times n_s$) are important for consistent synthesis. This simple yet effective approach to sampling better respects the relationship between features presented in the training data and results in generating 3D consistent subjects, despite utilizing 2D models. A similar idea for modelling the distribution of 3D brain MRI data via estimating the correlation in the latent space of a 2D slice VAE has recently been explored in \citep{volokitin2020modelling}.

\begin{figure}[!h]
    \centering
    \includegraphics[width=0.95\linewidth]{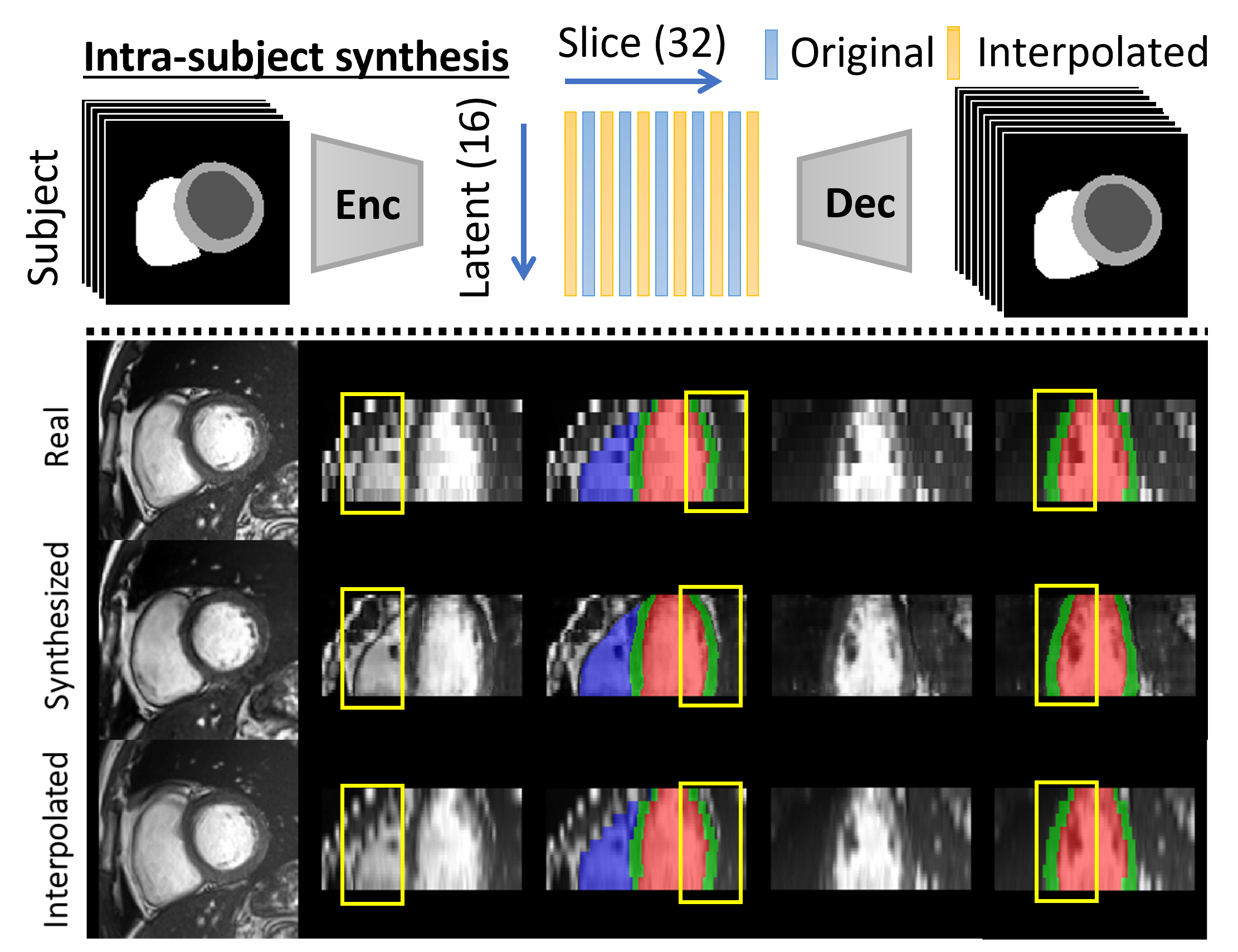}
    \caption{Intra-subject synthesis for increasing the resolution of the short-axis cardiac MR image stack by interpolating in latent space of GT labels for slices of the same subject and using them for synthesis. The original data contains around 8-10 slices per subject, and we create 32 slices using our intra-subject synthesis. The third row shows the result of image-based linear interpolation between slices.}
    \label{fig:my_label_2}
\end{figure}

\subsection{Data and Implementation}

To examine the ability of our inter-subject method to perform cross-vendor and cross-subject synthesis, we utilize cardiac MR images from a pair of subjects scanned using Siemens (vendor A) and Philips (vendor B) scanners provided by the M\&Ms-1 challenge \citep{MMs}. The disease information for each patient is required for our pathology synthesis experiment. For that purpose, we utilize ACDC challenge data \citep{ACDC} including normal cases (NOR) and three disease classes (heart dilated cardiomyopathy (DCM), hypertrophic cardiomyopathy (HCM), and abnormal right ventricle (DRV)). All 150 M\&Ms-1 and 100 ACDC subjects are resampled to $1.5 \times 1.5 mm$ in-plane resolution and cropped to $128 \times 128$ pixels around the heart using the provided ground truth labels. Percentile-based intensity normalization is applied as post-processing and the intensity range is mapped to the interval of -1 and 1. 

The input of the VAE model is a one-hot encoding version of the label map including three channels for cardiac classes right ventricle, left ventricle, myocardium, and background. The encoder part of the model includes four convolutional blocks with three convolutional layers each followed by batch normalization (BN) and LeakyReLU activation function. The encoded features are fed to four sequential fully connected layers to output the parameters of a Gaussian distribution over the latent representation. The decoder part of the model is comprised of four convolutional blocks each with one up-sampling layer followed by two convolutional layers with BN and LeakyReLU. The last additional block of the decoder includes one convolutional layer followed by BN and another convolution with four channel outputs and Softmax activation function. The VAE model is trained using a weighted combination of cross-entropy loss as the reconstruction loss and Kullback-Leibler divergence (KLD) with a weighting factor of $\beta$ for regularization of the latent space capacity \citep{higgins2016beta}. We experimentally identify the size of the latent vector ($n_z = 16$) and the weight of KLD ($\beta=15$) by inspecting the quality of the label reconstruction and the outcome of interpolation.

\begin{figure}[!ht]
    \centering
    \includegraphics[width=0.92\linewidth]{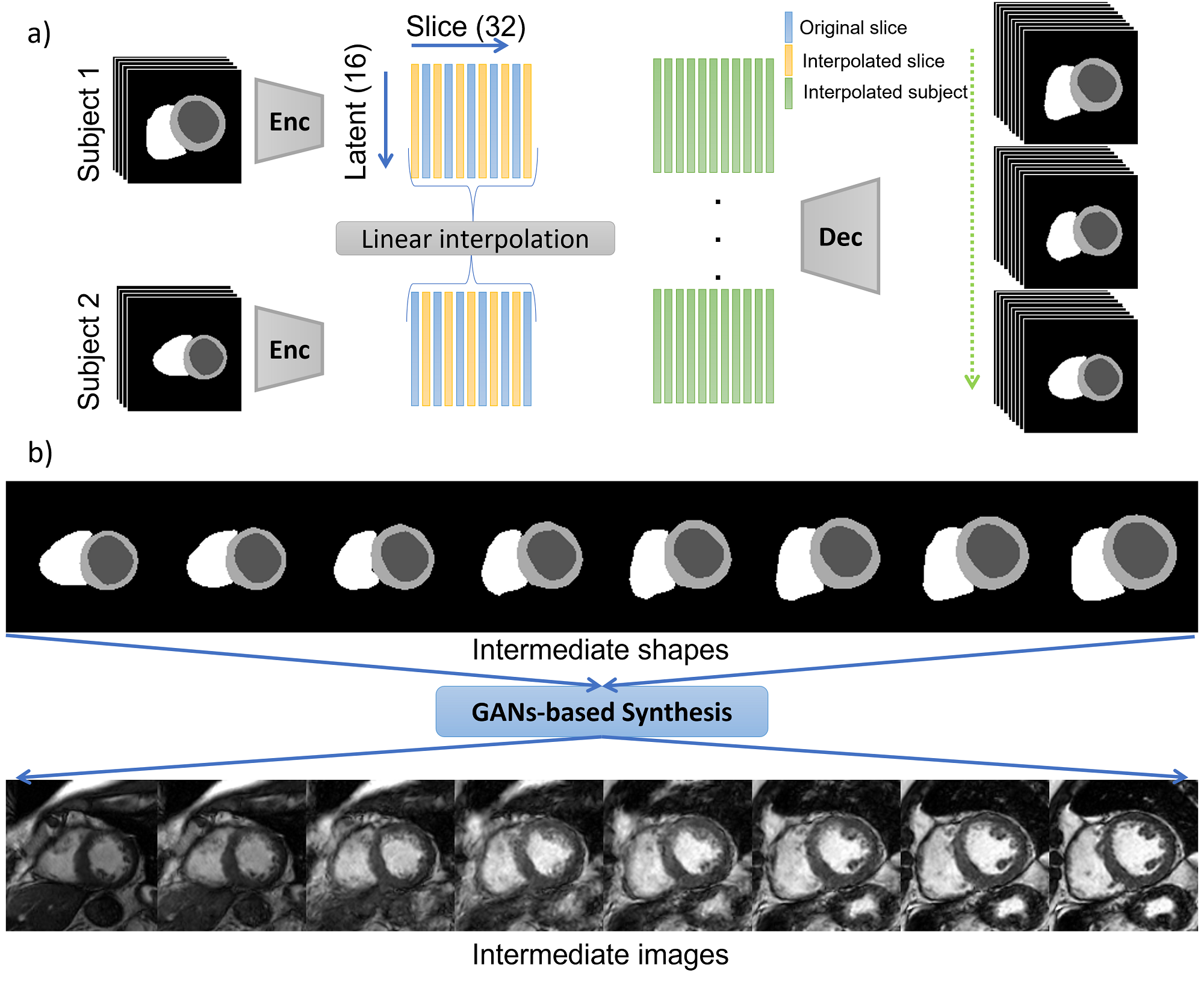}
    \caption{Inter-subject synthesis for generating intermediate shapes between two different heart geometries using linear interpolation between two subjects after equalizing the number of slices. The generated new labels are used for image synthesis.}
    \label{fig:my_label_3}
\end{figure}

\subsection{Usability of Synthetic Data}
\begin{figure}[!ht]
    \centering
    \includegraphics[width=0.95\linewidth]{ 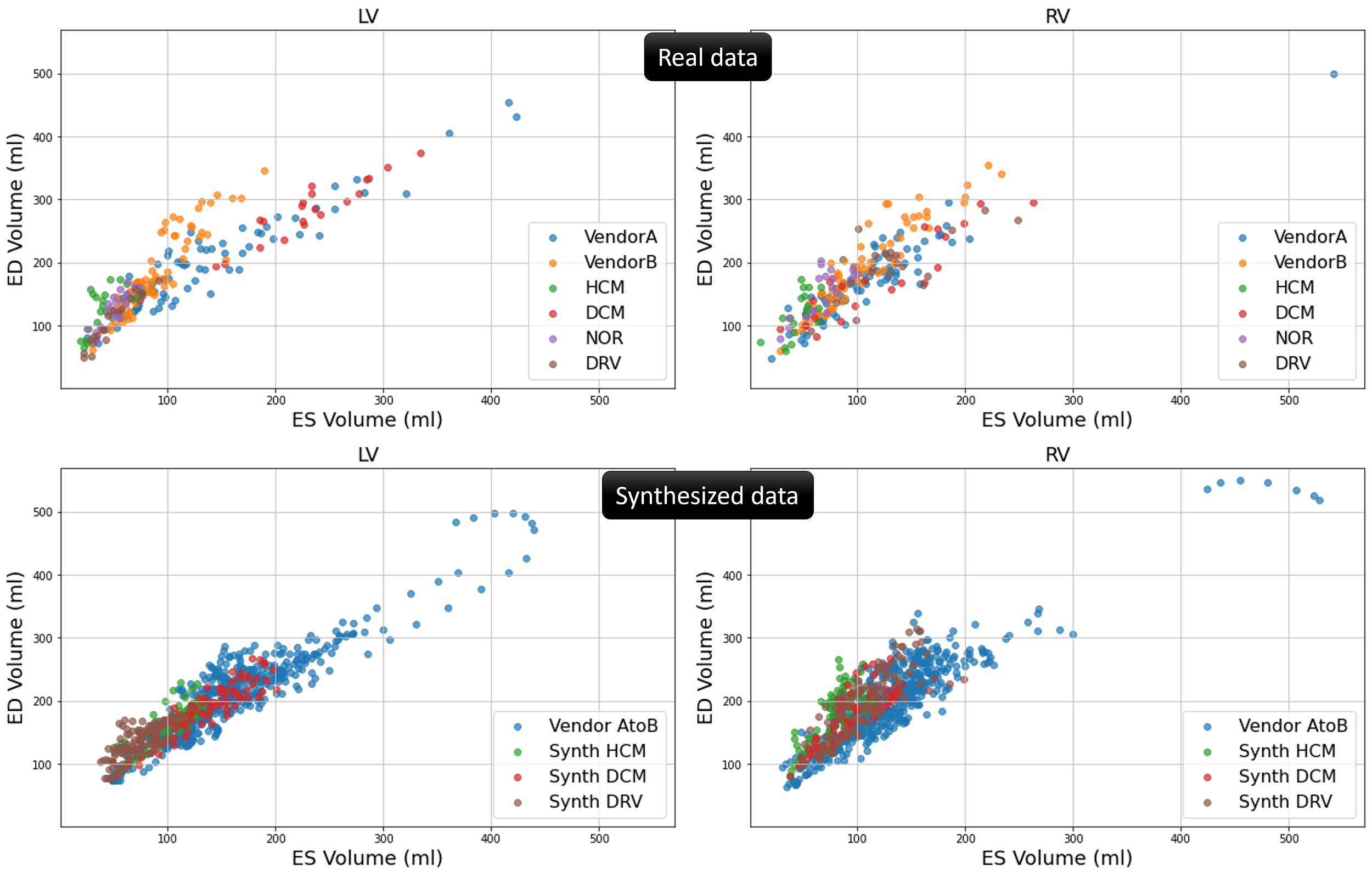}
    \caption{Distribution of calculated left- and right-ventricular volumes (LV and RV respectively) using the ground truth labels for end-diastolic (ED) and end-systolic (ES) phases of the heart for real and synthesized data. Vendor AtoB is the synthetic data for inter-subject synthesis between the data from M\&Ms-1 vendor A and vendor B subjects. The synthetic data for pathology synthesis between normal subjects (NOR) and the corresponding diseases such as Hypertrophic cardiomyopathy (HCM), Dilated Cardiomyopathy (DCM), and Right Ventricle Dilation (RV) are respectively denoted by Synth HCM, Synth DCM, and Synth RV}
    \label{fig:my_label_7}
\end{figure}

We generate synthetic data including five pathological versions of each NOR case from ACDC data. The Synth HCM data is generated by interpolating, in the latent space, between each NOR case and pseudo-pathological sample with characteristics of HCM subjects. The same applies to generating Synth DCM data and Synth RV data. Moreover, we employ the inter-subject synthesis approach to interpolate between the labels of subjects from vendor A and vendor B in the M\&Ms-1 challenge. By utilizing style images from both vendors during synthesis, we generate Vendor AtoB synthetic data which, as shown in Figure \ref{fig:my_label_5}, has characteristics of in-between vendor A and B data. To visualize the anatomical variation of the synthesized data in comparison with the real data, we calculate the end-diastolic (ED) and end-systolic (ES) volumes for the right ventricle and left ventricle blood pool using the ground truth labels. As can be seen from Figure \ref{fig:my_label_7}, there is a considerable similarity between the distribution of the synthesized data and the real data in terms of the calculated volumes.

We quantitatively evaluate the usefulness of the synthetic data for cardiac segmentation in the presence of pathologies and domain shift in cardiac MR image databases. Publicly available data typically suffer from the limited number of pathological cases, with rare diseases less likely to be represented well. Training with such data leads to models that struggle with generalization and adaptation to a wide variety of pathologies appearing among clinical cases. To tackle this, we utilize the synthetic data from our proposed inter-subject and pathology synthesis approaches in order to improve network generalization. To this end, we train six segmentation models with different combinations of real and synthesized data, to produce segmentation maps of three major heart structures - the left ventricle (LV), right ventricle (RV) blood pool and myocardium (MYO);

\begin{itemize}
    \item 1) \textbf{ACDC Real}: a model trained with 200 ED and ES real images acquired from the ACDC training set.
    \item 2) \textbf{ACDC Real + P. Synth.}: a model trained with 200 real ED and ES images from the ACDC training set and augmented with a total of 600 synthesized pathological cases for HCM, DCM, and RV (200 ED and ES images per pathology).
    \item 3) \textbf{ACDC M\&Ms Real}: a model trained with 200 real ACDC images and 300 M\&Ms-1 training images, acquired from vendors A and B (150 images each which include both ED and ES phases).
    \item 4) \textbf{ACDC M\&Ms Real + P. I. Synth.}: a model trained with real images from ACDC (200) and M\&Ms-1 (300) training data augmented with a combination of 600 synthesized pathological cases (as utilized for the augmentation of the \textbf{ACDC Real + P. Synth.} model) and 600 synthesized vendor AtoB images (I. Synth.).
    \item 5) \textbf{ACDC M\&Ms Real + P. Synth.}: a model trained with real images from ACDC (200) and M\&Ms-1 (300) training data augmented with 600 synthesized pathological cases.
    \item 6) \textbf{ACDC M\&Ms Real + I. Synth.}: a model trained with real images from ACDC (200) and M\&Ms-1 (300) training data augmented with 600 synthesized vendor AtoB images.
\end{itemize} 

To train the above segmentation models, we adapt a 2D nnU-Net \citep{nnunet} for a multi-class segmentation task with several modifications for improving the generalization and adaptation of the model to various data-sets used in this study, as proposed in \citep{full2020studying}. In particular, we replace the standard instance normalization layers of the baseline nnU-Net with batch normalization and introduce heavier data augmentation, besides the default transformations used within the nnU-Net pipeline. These include image scaling ($p=0.3$) with a scaling factor in the range of [0.7-1.4], random rotations within $\pm$60 degrees ($p=0.7$), random horizontal and vertical flips ($p=0.3$) and elastic transformations ($p=0.3$). Moreover, we apply intensity transformations in the form of gamma correction ($p=0.3$) with the gamma factor ranging within [0.5-1.6], additive brightness transformations ($p=0.3$) with the brightness factor varying within [0.7-1.3], multiplicative brightness ($p=0.3$) with a mean of 0 and standard deviation of 0.3 and the addition of Gaussian noise ($p=0.2$). During the pre-processing step, all data is normalized to an intensity range of [0-1], resampled to $1.5\times1.5mm$ in-plane resolution and center-cropped to $128\times128$ pixels around the heart, which is the same as the training patch size. 

We use a combination of Dice and cross-entropy loss for training, optimized using Adam for stochastic gradient descent, with an initial learning rate of 10$^{-4}$ and a weight decay of 3e$^{-5}$. During training, the learning rate is reduced by a factor of 5 if the validation loss has not improved by at least  $5\times 10^{-3}$ for 50 epochs. We train all models for a maximum of 1000 epochs, where early stopping is applied when the learning rate drops below 10$^{-6}$. Please note that we do not apply a cross-validation set-up during training and train all models once, utilizing all available images on four NVIDIA Titan Xp GPUs.

At inference time, we resample all images to $1.5\times1.5mm$ in-plane resolution and crop them around the heart area. Since the test images are typically of a larger field-of-view than those we use for training, and we cannot rely on the availability of labels, we apply a heart region detection network, proposed in \citep{al2021late}, responsible for obtaining the bounding box encompassing the whole heart. Before segmentation, the cropped images obtained using the predicted bounding boxes are post-processed to be of the size $128\times128$ pixels and normalized to the intensity range from 0 to 1. Finally, we perform a connected component analysis on the predicted
labels and remove all but the largest connected component per class.

We evaluate all six models on the hold-out data (completely unseen) from the M\&M-2 challenge with normal subjects (NOR) as well as various cardiac pathologies including Dilated Left Ventricle (DLV), Hypertrophic Cardiomyopathy (HCM), Congenital Arrhythmogenesis (ARR), Tetralogy of Fallot (FALL), Interatrial Communication (CIA), Tricuspidal Regurgitation (TRI), and Dilated Right Ventricle (DRV).
This allows us to study the generalization capability of both the baseline models ({ACDC Real} and {ACDC M\&Ms Real} trained with real images), as well as the models augmented with synthetic data generated in this study, to a wide array of moderate and severe pathological cases, some of which are not present in the training data. We report the segmentation performance in terms of Dice score (Dice) and Hausdorff Distance (HD), which are typically used as the main evaluation metrics in medical image segmentation challenges.

\section{Results}

\subsection{Intra- and Inter-subject Synthesis}

Figure \ref{fig:my_label_2} shows the results for slice interpolation in the intra-subject synthesis approach. The subject has originally nine slices and the synthesized version of the subject includes 32 slices. The effects on the through-plane resolution on both the image and on the ground truth labels can be observed from three orthogonal views of the cardiac MR image. Traditional image-based interpolation between slices can potentially result in a severe blurring effect due to the very large slice thickness of short-axis cardiac images. Additionally, the segmentation label masks are not properly preserved after image-based interpolation, especially the through-plane smoothness of the masks may not be achieved.


An example of inter-subject synthesis is given in Figure \ref{fig:my_label_3}. Note that the morphing from one shape to another is only shown for one mid-ventricular slice. The corresponding images are generated using the trained synthesis model.

\begin{figure}[!ht]
    \centering
    \includegraphics[width=0.95\linewidth]{ 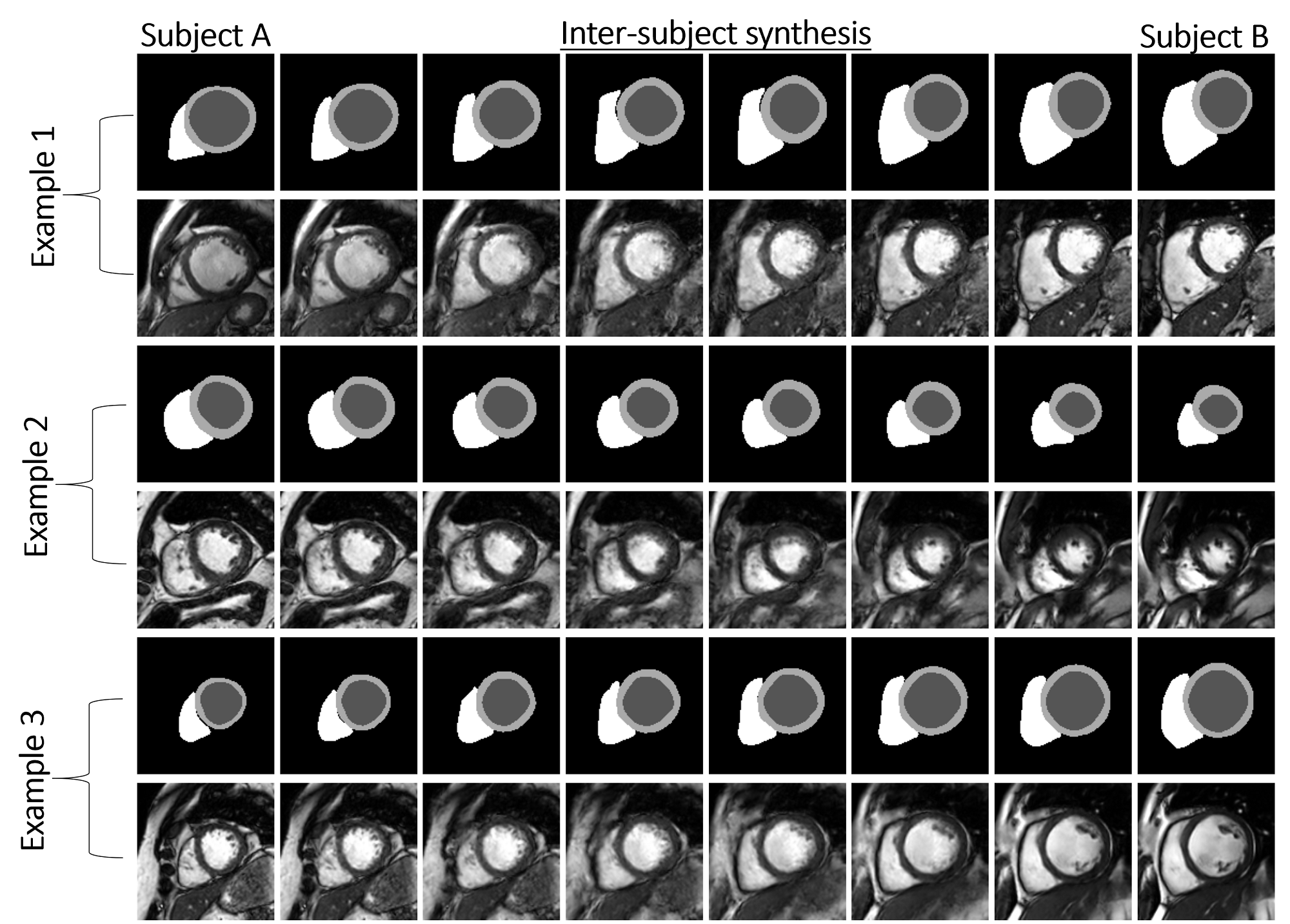}
    \caption{Inter-subject synthesis between three pairs of subjects chosen from the subset of vendor A and vendor B data. Subjects with similar heart shapes are chosen for example 2, whereas more dissimilar pairs are for 1 and 3. A smooth transition of the shape and appearance can also be observed for 1 and 3.}
    \label{fig:my_label_5}
\end{figure}

To examine the ability of our method for cross-vendor and cross-subject synthesis we choose pairs of subjects from Philips and Siemens subjects of M\&Ms database for inter-subject synthesis. We intend to show not only the morphing between two heart geometries but also to demonstrate the transition from one imaging characteristic to another one. Three examples for different levels of similarities between the heart geometries and image appearances are shown in Figure \ref{fig:my_label_5}. Example 1: for two rather similar heart shapes, example 2: from one heart with normal looking size to one with large left ventricle, and example 3: from another one with large left ventricle to one with large right ventricle.  A smooth transition between subjects from vendor A and another from vendor B can be observed from the results. 

\begin{figure}[!ht]
    \centering
    \includegraphics[width=0.91\linewidth]{ 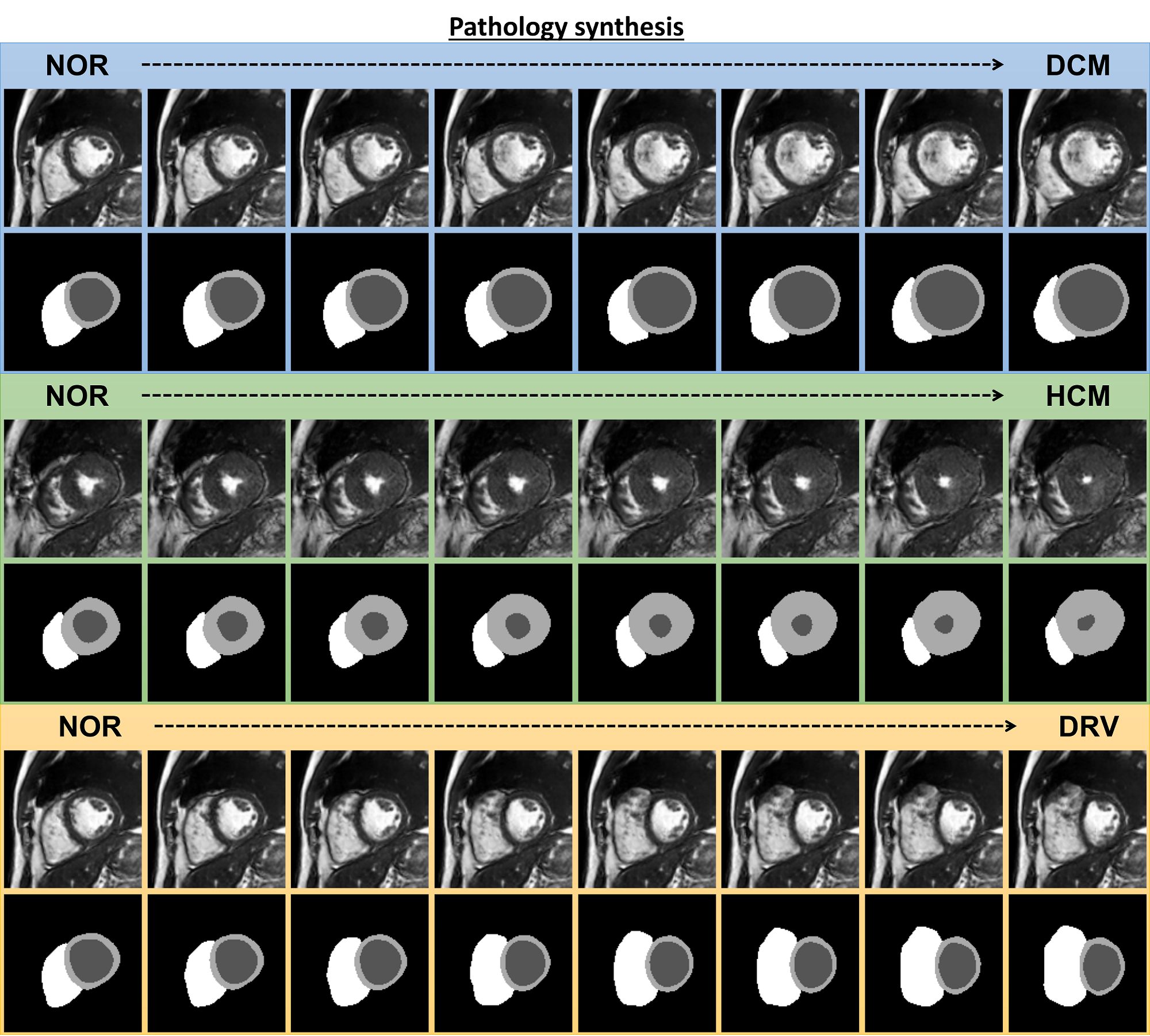}
    \caption{Pathology synthesis to generate the transition between a normal subject (NOR) to a target pathology such as dilated cardiomyopathy (DCM), hypertrophic cardiomyopathy (HCM) and dilated right ventricle (DRV). The effects of a disease on the heart geometry of a subject are respectively left ventricle dilation, myocardial thickening and right ventricle dilation.}
    \label{fig:my_label_6}
\end{figure}

\subsection{Pathology synthesis}

The results for pathology synthesis with three target heart diseases namely DCM, HCM, and DRV diseases are shown in Figure \ref{fig:my_label_6}. The characteristic of the particular heart disease is linearly added to the latent code of a normal subject (NOR). The heart shape characteristic of subjects with DCM, dilation of the left ventricle, is progressively appearing on the NOR subject through interpolation from left to right. The same is observed for thickening of the myocardium in the case of NOR to HCM and dilation of the right ventricle for NOR to DRV. Note that in pathology synthesis, in contrast to inter-subject synthesis, the identity of the NOR subject is not changing while the disease features are manifested on the geometry of the subject’s heart and the image appearance stays the same. The identity of a normal subject in this context refers to the subject-specific heart shape, image appearance/contrast, and the positioning of the heart with respect to other surrounding organs. When we perform inter-subject synthesis between subjects A and B, all of these features are changed for the intermediate generated subjects due to the alteration of the input style image. Therefore, the identity of the subject is not preserved. However, during pathology synthesis, we use the NOR subject as the style, which means that only the heart shape characteristics are deformed by the VAE in a way that produces a diseased heart shape for the generated subjects. Since the image appearance/contrast and other organs remain unchanged, we can say that the identity of the NOR subject is preserved and only the diseased condition is generated. Interestingly, the detailed structures of the papillary muscles and myocardial trabeculations inside the left and right ventricular blood pool are generated despite not being present in the ground truth labels. 

\begin{figure}[!ht]
    \centering
    \includegraphics[width=0.6\linewidth]{ 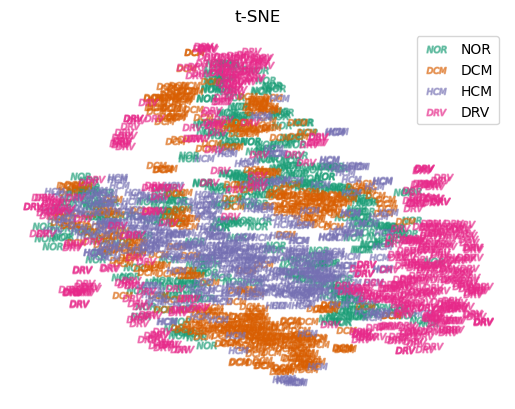}
    \caption{The t-SNE embedding to visualize the latent representation of each encoded slice for all subjects, with slices of normal subjects (NOR) depicted in green, dilated cardiomyopathy (DCM) in orange, hyperthrophic cardiomyopathy (HCM) in purple, and dilated right ventricle (DRV) in pink. }
    \label{fig:tsne}
\end{figure}

To assess our initial assumption that the ground truth labels for subjects with the same pathology tend to cluster together in a neighboring location in the latent space, we employ t-SNE \cite{van2008visualizing} to analyze the latent representation of each encoded slice for all subjects. The results of this analysis are presented in Figure \ref{fig:tsne}, where we can observe distinct clusters in the embedding space for different heart diseases.

\subsection{Modeling the slice relationship}

\begin{figure}[!ht]
    \centering
    \includegraphics[width=0.87\linewidth]{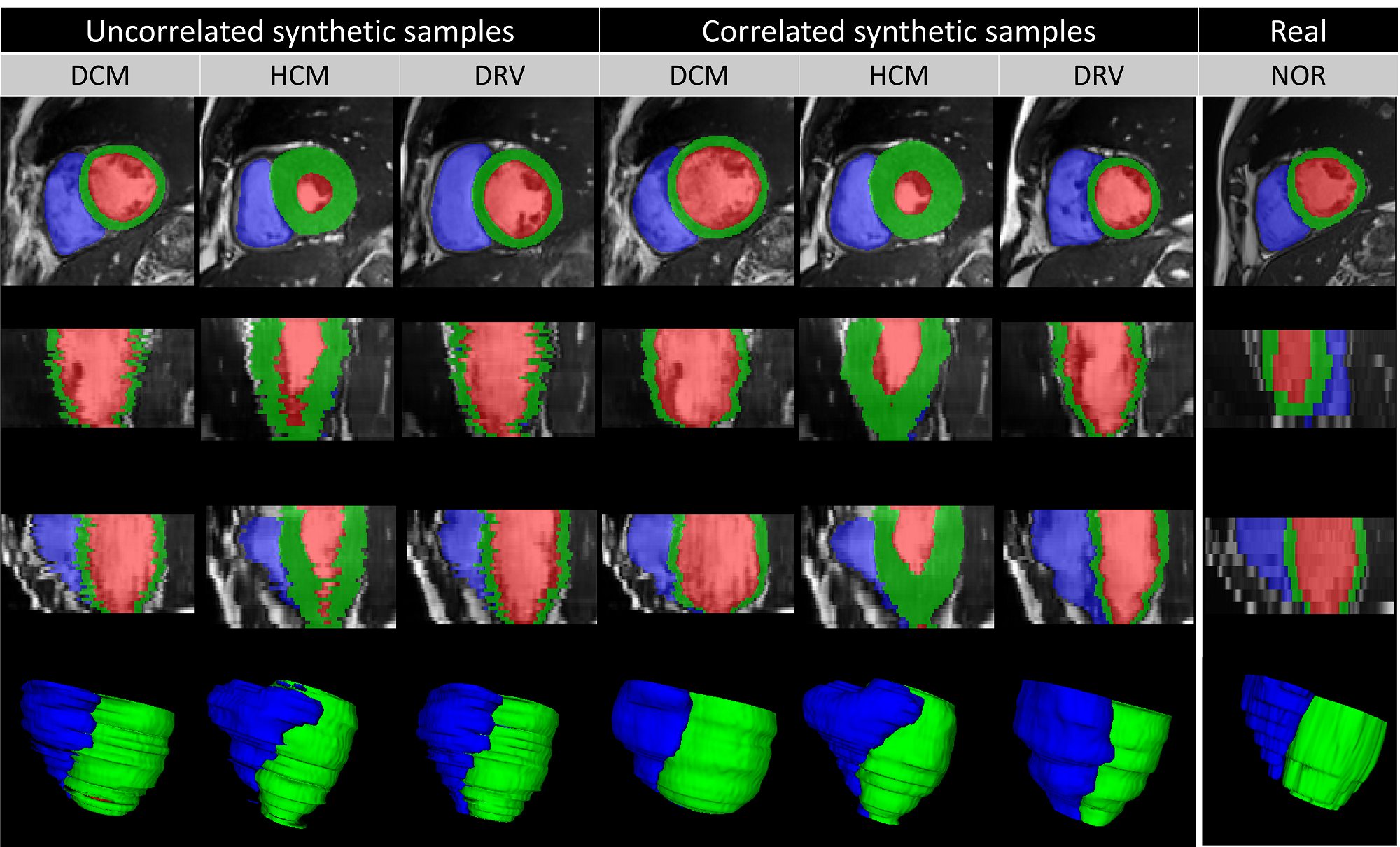}
    \caption{Three-dimensional rendering of the labels for uncorrelated and correlated synthesis for different cases of pathology synthesis. The first three columns show the uncorrelated slices and the impact of the inconsistency of the anatomy in the perpendicular views of the short axis slices while the second three columns show the positive effects of correlating samples on reducing the inconsistency and irregularity of the consecutive slices. The last column shows one real example.}
    \label{fig:my_label_8}
\end{figure}

Our proposed 2D model synthesizes images slice-by-slice with high visual fidelity and realism. However, the synthetic subject that is composed of stacking multiple 2D slices is not generated coherently by the network when we look at the generated slices from perpendicular directions. The reason is that random samples in the latent space contain no information about the relationship between different slices of one subject, i.e. generated slices are uncorrelated. Synthesis examples with target pathologies and the positive effects of the proposed slice correlation on generating 3D consistent subject are shown in Figure \ref{fig:my_label_8} with a three-dimensional rendering of the synthesized labels. The irregularities in the slice direction are substantially reduced for the correlated slices for synthesizing different pathological cases. We notice that some real images may originally be hampered by slice misalignment artifacts and our correlated sampling cannot reduce this artifact.

\subsection{Usability of Synthetic Data}

\begin{figure}[!ht]
    \centering
    \includegraphics[width=0.99\linewidth]{ 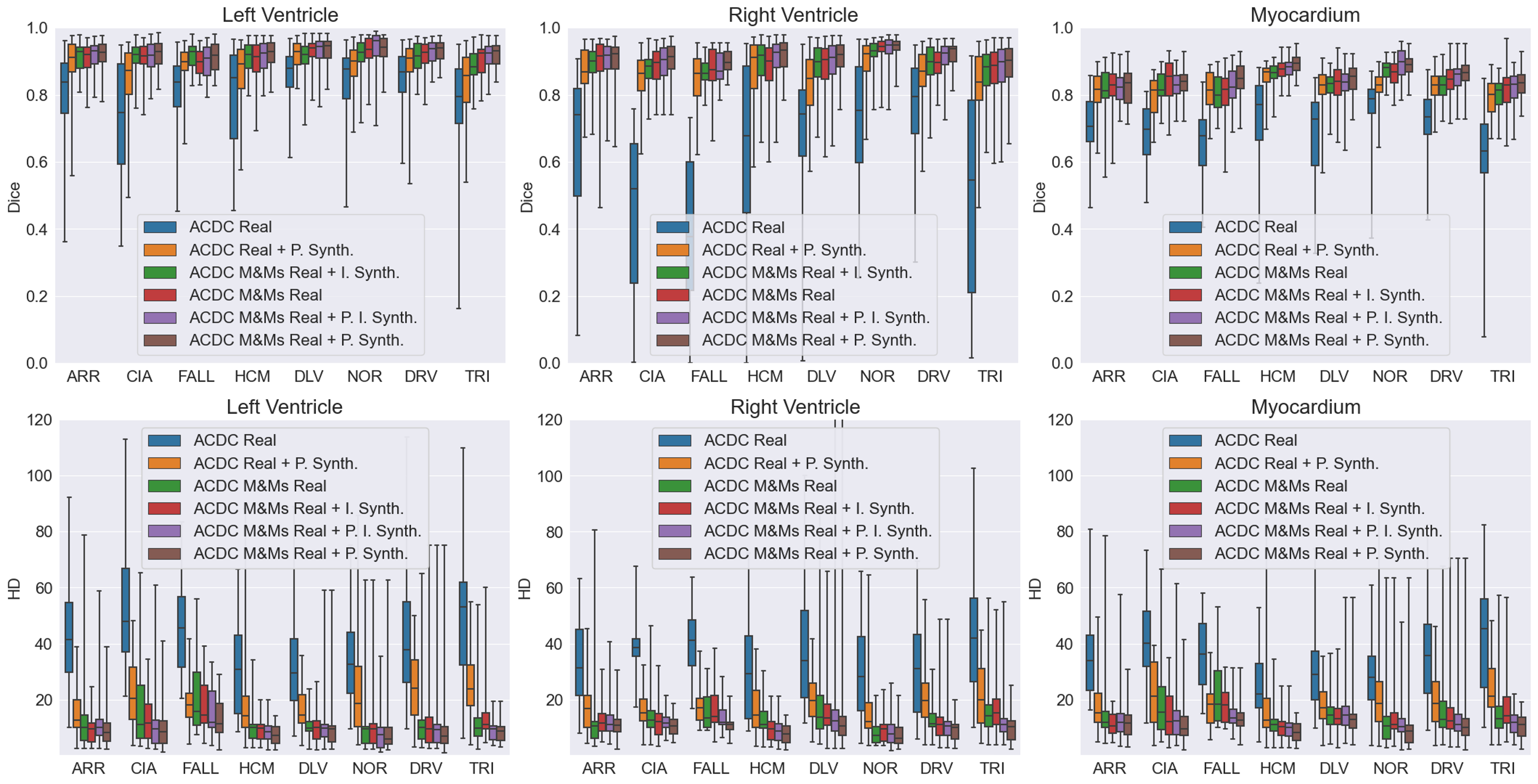}
    \caption{Dice scores and Hausdorff Distance (HD) performance on the unseen data from M\&Ms-2 challenge with different cardiac pathologies. }
    \label{fig:my_label_10}
\end{figure}

Figure \ref{fig:my_label_10} shows the performance of four segmentation models in terms of Dice and Hausdorff distance (HD) for left ventricle, right ventricle and myocardium segmentation. A significant drop in the performance observed for the ACDC real model (trained using real ACDC data only) suggests a substantial domain shift (distributional shift) between ACDC and M\&Ms-2 images, mostly due to the acquisition hardware, protocols, and presence of unseen heart pathologies. While the addition of M\&Ms-1 images to the training (ACDC M\&Ms model) helps significantly with tackling this domain shift and improving generalization, we also note that the addition of the synthesized data with pathological characteristics substantially improves the segmentation performance and the robustness of the model (ACDC Real + P. Synth.) across all cardiac diseases. This indicates that the pathology synthesis approach can generate realistic images with relevant pathological diversity for training a cardiac segmentation network that generalizes well to even unseen diseases during training. Moreover, it also suggests that synthetic images generated in this study serve as a realistic substitute for real MR images when met with limitations in acquiring more data from the target domain.

While augmentation with inter-subject synthesis data (ACDC M\&Ms Real + I. Synth.) slightly improves the baseline performance in the of reduction in HD score, the improvement in Dice score is marginal and even decreased for some pathologies compared to the model trained with all real images. We hypothesize that this might be due to the generation of heart shapes that may not necessarily be similar to the pathological cases in the test set. Additional benefits are observed when further adding the pathology synthesis data (ACDC M\&Ms Real + P. I. Synth). Augmentation with pathology synthesis data (ACDC M\&Ms Real + P. Synth.), however, is found to be crucial for significantly improving the segmentation performance on unseen heart pathologies. We attribute the superiority of pathology synthesis to the fact that the method is tailored to generate subjects with pathological characteristics which helped with better generalization and performance on the testing data containing various pathologies, resulting in the best performance across all three cardiac tissues and varying pathologies. In other words, the best performance in Dice and HD across all cardiac diseases is obtained by the model trained using only pathology synthesis, indicating the usefulness of the generated images for clinical tasks. We particularly note a significant improvement in HD scores, primarily due to outlier reduction, as well as a decrease in the number of over-segmented and under-segmented tissues. Under-segmentation due to diseased tissue and occlusion is a common problem observed for baseline models, which can be tackled by the augmentation approach proposed in this study, allowing the network to learn from a much higher and diverse pool of examples than just relying on carefully curated and limited real data. We hypothesize that synthesized images contribute to a higher variation in heart tissue shape and appearance, particularly those undergoing changes due to the presence of pathology, which in turn helps with network regularization and generalization. 
 

\section{Discussion and Conclusion}

This study investigated an approach for realistic cardiac magnetic resonance image synthesis with target heart pathologies by separating the task into label deformation using a VAE and image generation using a label-conditional GAN. We introduced intra-subject synthesis to increase the through-plane resolution of short-axis images and to equalize the number of slices across all subjects. The inter-subject synthesis was designed to perform cross-subject and cross-vendor synthesis by generating subjects that have intermediate heart geometries and appearances between two dissimilar subjects scanned using different vendors. Furthermore, the pathology synthesis was proposed to generate subjects with heart characteristics of a particular disease through sampling in the latent space with statistics of a target pathology and performing linear interpolation between a normal subject and a pseudo-pathological sample in the latent space of the trained VAE.

To tackle one of the important challenges of 3D medical image synthesis, we demonstrated that modelling the correlation between slices in the latent space can be a simple yet effective way to generate consistent 3D subjects from 2D models. 

Visualizations of the synthesized images and the distribution of the left and right ventricular volumes on the synthesized data showed encouraging results. Moreover, we devised experiments to quantitatively evaluate the usability of the synthetic data for the development of a generalizable deep learning segmentation network.  We found that both generated images by inter-subject and pathology synthesis are extremely useful in improving the generalization and robustness of a deep-learning segmentation model in a challenging clinical environment. The methods proposed in this study could be extended for other applications in medical image synthesis such as brain MR image generation and simulation of lesion progression.

A limitation of our study is the lack of a measure for assessing the 3D consistency of the synthesized subjects. Moreover, the result of the intra-subject synthesis is not quantitatively evaluated on its own as an independent approach for increasing the resolution in the slice direction and the effect on segmentation, apart from qualitative results shown in Figure \ref{fig:my_label_2}. Note that this is a necessary step for inter-subject and pathology synthesis to equalize the number of slices for each subject. Another limitation is that we in fact evaluate the usefulness of the synthetic images for data augmentation, while quantifying the level of image realism for each synthetic subject remains to be explored. For instance, when interpolating from a normal subject to a pathological sample, it is likely, but not guaranteed that the intermediate heart changes reflect what truly happens in disease progression. Similarly, when interpolation between two subjects, not all intermediate shapes reflect what can be found in real life. Despite all that, we examine the benefit of the synthetic images for training a deep learning model which indicates the usefulness of the generated images for data augmentation. 

In conclusion, we demonstrated that our approach could provide a solution to diversify and enrich an available database of cardiac MR images, resulting in significant improvements in model performance and generalization for cardiac segmentation of subjects with unseen heart diseases.


\acks{This research is a part of the OpenGTN project, supported by the European Union in the Marie Curie Innovative Training Networks (ITN) fellowship program under project No. 76446}

%

\ethics{The work follows appropriate ethical standards in conducting research and writing the manuscript, following all applicable laws and regulations regarding the treatment of animals or human subjects.}


\coi{We declare we do not have conflicts of interest.}

\bibliography{sample}


\newpage

\appendix

\section{Cholesky decomposition and correlated samples}

In order to simulate correlated variables with a given covariance matrix ($C$), Cholesky matrix decomposition is used in this study. The Cholesky matrix decomposition is a factorization of a positive-definite symmetric matrix into a product of a lower and upper triangular matrix, $L$ and $L^T$, respectively.
\begin{equation}
 C=L L^T  
\end{equation}
Assuming an uncorrelated random sample $X$ with unit covariance matrix of $E(XX^T )=I$, a new random vector can be computed as $Y=LX$ that its covariance matrix is derived:
\begin{equation}
E(YY^T )= E(LX(LX)^T )= E(LXX^T L^T )=LE(XX^T ) L^T=LIL^T=LL^T=C
\end{equation}
Note that the expectation is a linear operator; $E(cX)=cE(X)$.

\section{Correlating and generating sample with pathology characteristics}

For generating a subject with pathological characteristics, a random sample is drown using a truncated normal distribution parameterized by the statistics of the desired pathology, e.g. mean, standard deviation, minimum, and maximum estimated on all subjects with hypertrophic dilated cardiomyopathy (HCM); namely pseudo-pathological sample  $x_pHCM$. These statistics are calculated on the mean of the posterior distribution of features estimated by the encoder part of the VAE.  The following steps are followed to correlate the elements of this pseudo-pathological sample cross slice direction and latent dimension:

\begin{itemize}
    \item Estimate correlation coefficient between latent dimensions across all subjects with desired pathology using Kendall rank correlation coefficient method; $Corr_{zHCM}$ with size ($n_z\times n_z$) where $n_z$  is the size of the latent vector ($n_z=16$)
    \item Calculate the lower triangular matrix $L$ using Cholesky decomposition; $L_{zHCM}$
    \item Correlate the latent dimensions of the pseudo pathological sample across the element of latent vector given above formula; $y_{pzHCM} = {L_{zHCM}} {x_{pHCM}}$
    \item Estimate the correlation coefficient between slices of the target normal subject (NOR) we wish to use for interpolation; $Corr_{sNOR}$ with size ($n_s\times n_s$) where $n_s$  is the number of slices ($n_s=32$)
    \item Calculate the lower triangular matrix $L$ using Cholesky decomposition; $L_{sNOR}$
    \item Correlate the latent dimensions of the pseudo random sample cross slices given above formula; $z_{pzsHCM} = {L_{sNOR}} {y_{pzHCM}}$
    \item Linearly interpolate between $z_{NOR}$ and $z_{pzsHCM}$ in the latent space
    \item Reconstruct slices-by-slice the interpolated samples using the decoder part of the 2D VAE
    \item Compose 3D volume from synthesized 2D slices
\end{itemize}

\begin{figure}[!ht]
    \centering
    \includegraphics[width=0.95\linewidth]{ 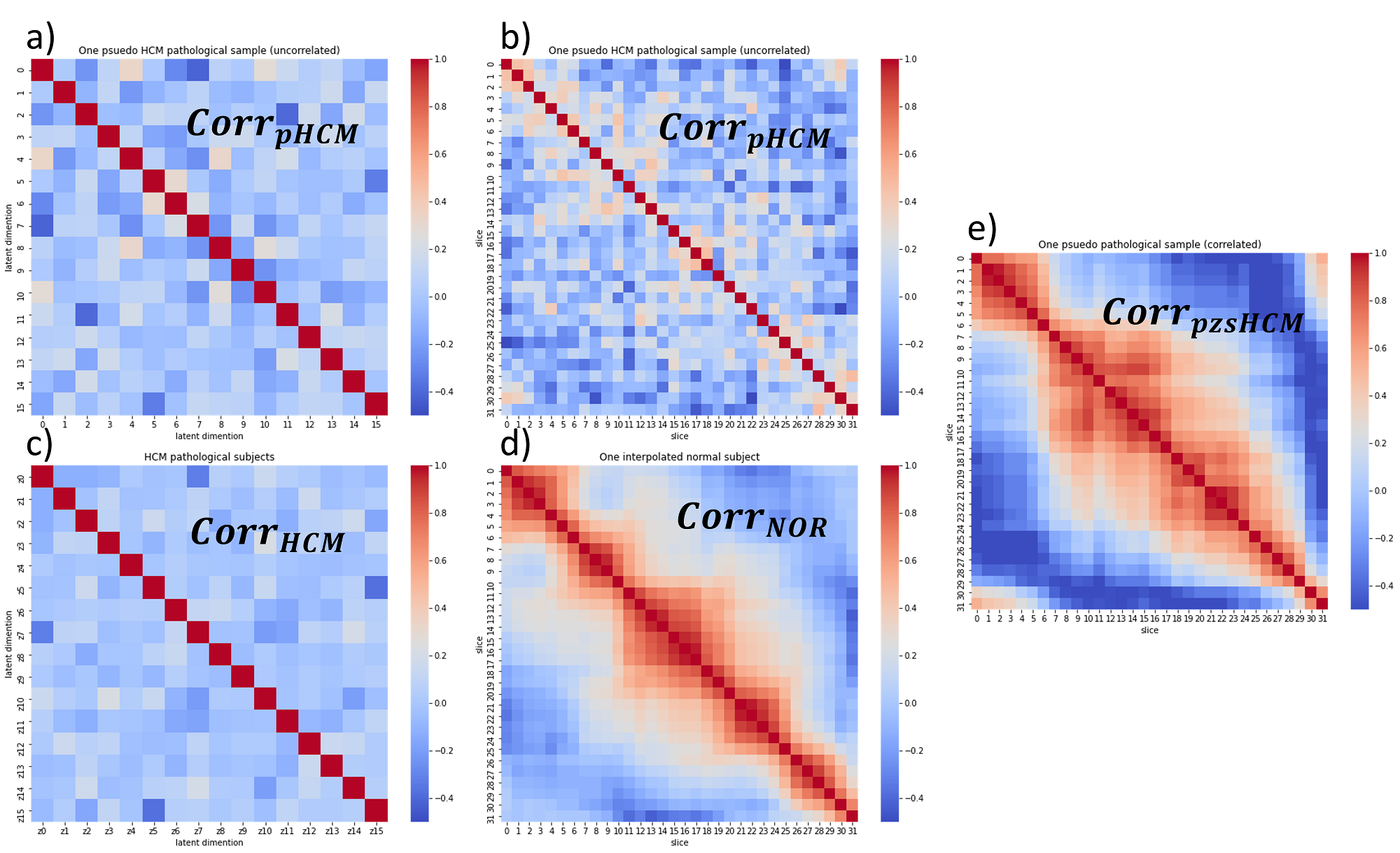}
    \caption{Correlation coefficient matrix for a) uncorrelated pseudo-HCM sample across latent dimensions and b) across slices, c) all HCM subjects across latent dimensions, d) one normal subject across slices, and e) the correlated pseudo pathological sample calculated using the Cholesky decomposition.}
    \label{fig:my_label_9}
\end{figure}

The correlation coefficient matrix for all above mentioned steps is shown in Figure \ref{fig:my_label_9}. Correlating latent dimensions found to be as important as correlating slices of subject for generating coherent slices with smoothly changing features.

\end{document}